\documentclass[aps,pra,twocolumn,superscriptaddress,notitlepage]{revtex4-2}
\usepackage{tikz}
\usetikzlibrary{calc}
\usetikzlibrary{shapes.geometric, arrows}
\usepackage{eufrak}
\usepackage{caption}
\usepackage{enumitem}
\usepackage{mathtools}
\usepackage{graphicx}
\usetikzlibrary{positioning}
\usepackage{bm}
\usepackage{setspace}
\usepackage{xcolor}
\usepackage[justification=justified,singlelinecheck=false]{caption}
\newcommand{\bematrix}{\left(\begin{matrix}}
\newcommand{\ematrix}{\end{matrix}\right)}

\usepackage{ulem} 
\usepackage[caption=false]{subfig}
\usepackage{dcolumn}
\usepackage{amsmath,amssymb}
\usepackage{bm}
\usepackage{bbm}
\usepackage{overpic}
\usepackage{latexsym}
\usepackage{color}
\usepackage[english]{babel}
\usepackage{latexsym}
\usepackage{psfrag,graphicx}
\usepackage{epsf}
\usepackage{amsmath}
\usepackage{amssymb}
\usepackage{amsfonts}
\usepackage{natbib}
\usepackage{multirow} 
\usepackage{appendix}
\usepackage{verbatim}
\usepackage{enumitem}
\usepackage{dsfont}

\usepackage{float}
\usepackage{tikz}

\definecolor{mygrey}{gray}{0.35}
\definecolor{myblue}{rgb}{0.2,0.2,0.8}
\definecolor{myzard}{cmyk}{0,0,0.05,0}
\definecolor{mywhite}{rgb}{1,1,1}
\definecolor{myred}{rgb}{0.9,0.1,0.}
\usepackage[colorlinks=true,citecolor=myblue,linkcolor=myblue,urlcolor=myblue]{hyperref}

\usepackage[makeroom]{cancel}

\newenvironment{proof-of}[1]{\medskip\noindent\textbf{Proof of {#1}.}}{\hfill$\blacksquare$\medskip}
\newcommand{\ket}[1]{\left\vert#1\right\rangle}
\newcommand{\bra}[1]{\left\langle#1\right\vert}

\usepackage{colortbl}
\usepackage{xcolor}

\usepackage{tcolorbox}

\definecolor{lightgray}{gray}{0.9}

\makeatletter
\long\def\@makecaption#1#2{%
  \par\begingroup
  \small
  \@parboxrestore
  \rightskip=0pt
  \leftskip=0pt
  \parfillskip=0pt
  \noindent
  #1.~#2\par
  \endgroup}
\makeatother

\begin{document}

\title{Cavity-Free Distributed Quantum Computing with Rydberg Ensembles via Collective Enhancement}

\author{Muhammad Ali Shahbaz}
\affiliation{Department of Physics, University of Nevada, 89516, Reno, USA}

\author{Aman Ullah}
\email{aullah@unr.edu}
\affiliation{Department of Physics, University of Nevada, 89516, Reno, USA}

\begin{abstract}
\noindent We present a complete protocol for cavity-free quantum networking based
on collective enhancement in Rydberg atom ensembles. The scheme combines
Rydberg blockade, collectively enhanced light--matter coupling, and
phase-matched directional emission to remove the need for optical
cavities while retaining efficiencies comparable to cavity-assisted
interfaces. The protocol proceeds in three steps: (i)~local
control--ensemble entanglement generated by Rydberg blockade with gate
fidelity $F_{\mathrm{gate}}\approx 99.93\%$; (ii)~atom--photon
conversion through Raman emission from an oblate spheroidal ensemble,
yielding directional emission efficiency $\eta_{\mathrm{dir}}\approx 73\%$
and single-node efficiency $\eta_{\mathrm{node}}\approx 40\%$; and
(iii)~remote atom--atom entanglement via Hong--Ou--Mandel interference,
producing Bell states with fidelity $F>97.5\%$. Incorporating quantum
memories allows up to $M\approx 100$ retry attempts within a coherence
time $T_2>100\,\mu\mathrm{s}$, enabling entanglement generation rates
of approximately $4\,\mathrm{kHz}$ over a 20~km separation.
Collectively enhanced Rydberg ensembles thus provide a practical,
cavity-free interface for scalable distributed quantum computing and
secure quantum communication.
\end{abstract}

\date{\today}

\maketitle

\section{Introduction}
Quantum networks distribute entanglement between distant nodes, enabling secure communication, and distributed quantum computing~\cite{ekert1991quantum,cirac1999distributed,degen2017quantum, kimble2008quantum,wehner2018quantum,knaut2024entanglement,main2025distributed}. Significant progress, both experimental and theoretical, toward realizing such networks has been achieved across different physical platforms, including trapped ions~\cite{bruzewicz2019trapped,blatt2012quantum,drmota2023robust,pompili2021realization}, nitrogen-vacancy centers in diamond~\cite{nemoto2014photonic,yao2012scalable,maurer2012room}, and neutral atoms~\cite{saffman2019quantum,henriet2020quantum,covey2023quantum, knorzer2025distributed}. Among these, neutral atoms are particularly attractive due to their long coherence times, strong and controllable Rydberg-mediated interactions, and natural compatibility with photonic interfaces~\cite{xia2015randomized,wang2015single,dudin2012single,xu2021fast}.

The Rydberg blockade mechanism—where a single Rydberg excitation suppresses additional excitations within a blockade radius via strong van der Waals interactions—provides a powerful tool for quantum information processing. This effect has enabled high-fidelity two-qubit gates~\cite{evered2023high,ma2023high,isenhower2010demonstration,levine2019parallel,graham2019rydberg}, multi-particle entangled states~\cite{wilk2010entanglement,zeiher2016many,bluvstein2024logical}, and programmable quantum simulators~\cite{bernien2017probing,scholl2021quantum}. Within the same Rydberg framework, entanglement between a control atom and an ensemble of $N$ atoms has been explored as a resource for quantum networking, leveraging the ensemble's collective enhancement for efficient light-matter interfaces~\cite{saffman2005entangling,dudin2012strongly,dudin2012observation,weber2015mesoscopic,spong2021collectively}. Separately, atom-photon coupling and heralded photonic entanglement have been demonstrated in cavity-based systems~\cite{thompson2013cavity,togan2010quantum}. Cavities, while effective, impose narrow bandwidths and introduce experimental complexity that hinders scalability.  However, despite progress in these individual components, a unified cavity-free protocol that integrates deterministic local entanglement, efficient photon generation, and heralded remote entanglement into a complete quantum network architecture remains unrealized.

\begin{figure*}[t]
\includegraphics[width=\textwidth,keepaspectratio]{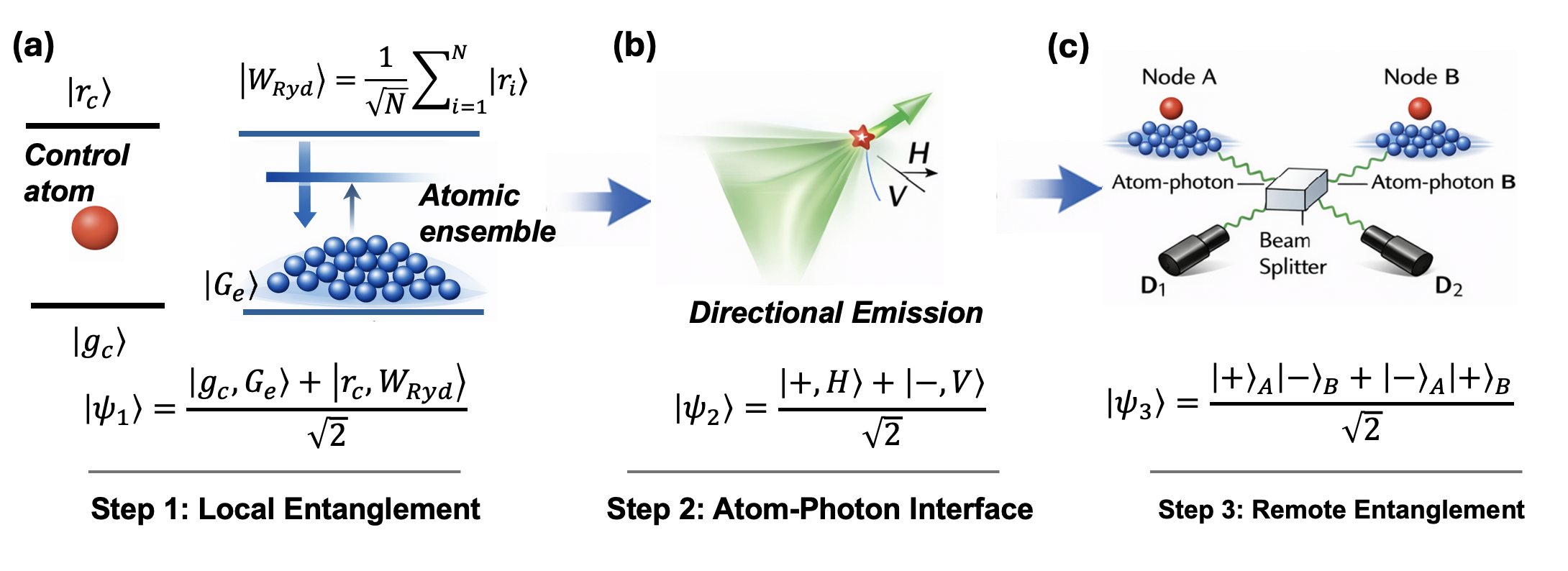}
\caption{ \textbf{Three-step protocol for cavity-free quantum networking with Rydberg atoms.}
\textbf{(a)-} A control atom (red) and atomic ensemble are entangled via Rydberg blockade. Energy levels show the control atom states $|g_c\rangle$ (ground) and $|r_c\rangle$ (Rydberg), and ensemble states $|G_e\rangle$ (collective ground state) and $|W_{\mathrm{Ryd}}\rangle$ (collective Rydberg excitation), generating the entangled state $|\psi_1\rangle$.
\textbf{(b)-} Raman transitions map the collective excitation to a directional photon (green cone). State-selective coupling to Rydberg magnetic sublevels encodes quantum information in photon polarizations $|H\rangle$ and $|V\rangle$, creating the atom-photon state $|\psi_2\rangle$.
\textbf{(c)-} Constructive interference at a beam splitter with heralding detectors (D$_1$, D$_2$) generates the remote Bell state $|\psi_3\rangle$.}
\label{fig:protocol_overview}
\end{figure*}

In this work, we have presented a comprehensive three-step protocol (Fig.~\ref{fig:protocol_overview}) that provides a complete cavity-free quantum network architecture. In the first step, the protocol  exploits Rydberg blockade and collective coupling enhancement to generate high-fidelity entanglement $\ket{\psi_1}$ between a control atom and an ensemble of $N $ atoms (see figure~\ref{fig:protocol_overview}a). A control atom is prepared in a coherent superposition of its ground and Rydberg states, while strong van der Waals interactions confine the ensemble to a single collective Rydberg excitation shared across all atoms. This collective excitation yields an enhanced effective Rabi frequency $\Omega_{\mathrm{eff}} = \sqrt{N}\Omega$, where $\Omega$ is the single-atom Rabi frequency, enabling fast control while Rydberg blockade restricts dynamics to the single-excitation manifold~\cite{ebert2014atomic,saffman2010quantum}. 
In the second stage, the locally generated entanglement is converted into atom–photon entanglement, $\ket{\psi_2}$, through Raman transitions that map the collective excitation onto a propagating optical mode (see figure~\ref{fig:protocol_overview}b). By imprinting a spatial phase gradient across the ensemble, phase-matched, directional photon emission is achieved, allowing efficient coupling into an optical fiber. The Raman process further enables state-selective mapping between Rydberg magnetic sublevels and photon polarization, thereby encoding the photonic qubit in polarization states $|H\rangle$ and $|V\rangle$ that remain entangled with the control atom. In the final step, photons emitted from two independent nodes are interfered on a beam splitter~\cite{hong1987measurement}, where Hong--Ou--Mandel interference and heralded single-photon detection~\cite{duan2001long} probabilistically project the remote atoms into an entangled Bell state, $\ket{\psi_3}$ (see figure~\ref{fig:protocol_overview}c).
The combination of directional emission and polarization encoding supports high collection efficiencies, making the system well suited for remote entanglement generation. Together, these steps establish a deterministic local interface between atomic and photonic qubits and enable the heralded generation of remote atom--atom entanglement, providing a scalable building block for quantum networking.

The present manuscript is organized as follows. Section~\ref{local-entanglement} describes the generation of local entanglement between a control atom and a Rydberg ensemble via the blockade mechanism. Section~\ref{atom_photon} presents the atom--photon interface, covering the Raman mapping process, the directional emission geometry, and the polarization encoding of the photonic qubit. Section~\ref{sec:step3} analyzes the remote entanglement step, in which photons from two nodes interfere on a beamsplitter and coincidence detection heralds a Bell state shared between the nodes. Section~\ref{prot-eff} evaluates the overall protocol efficiency, including the role of quantum memory in enabling repeated attempts and the resulting entanglement generation rate of ${\approx}4\,\mathrm{kHz}$ at 20~km separation. Finally, we conclude that the elimination of optical cavities, combined with collective enhancement in Rydberg ensembles, offers a scalable and experimentally feasible pathway toward metropolitan-scale distributed quantum computing.

\section{Local Entanglement}\label{local-entanglement}
The local entanglement between a control atom and a collective Rydberg excitation within an atomic ensemble (Fig.~\ref{fig:protocol_overview}a) is generated via the Rydberg
blockade mechanism. The system consists of a single control atom with ground state $\ket{g_c}$ and Rydberg state $\ket{r_c}$, and an ensemble of $N \approx 1000$ atoms — chosen to balance collective enhancement ($\Omega_{\mathrm{eff}} = \sqrt{N}\,\Omega$, enabling nanosecond gates) against sensitivity to atom-number fluctuations and multi-level leakage that grow with $N$ (discussed below) — each with ground state $\ket{g}$ and Rydberg state $\ket{r}$. The ensemble is initially prepared in the collective ground state $\ket{G_e} = \bigotimes_{i=1}^{N}\ket{g_i}$, where $\ket{g_i}$ denotes the ground state of atom $i$.

Rydberg--Rydberg interactions are governed by the repulsive ($C_6 > 0$) van der Waals potential $V(r) = C_6/r^6$, where $C_6$ is the dispersion coefficient for the target Rydberg state. The blockade radius $r_b = \left(C_6/\hbar\Omega\right)^{1/6}$ is defined by the condition $V(r_b) = \hbar\Omega$: for $r < r_b$, the interaction energy exceeds the single-atom Rabi frequency and multiple simultaneous Rydberg excitations become energetically inaccessible. This blockade confines the ensemble dynamics to the single-excitation manifold spanned by $\ket{G_e}$ and the singly-excited states $\ket{r_i} \equiv \ket{g_1\cdots r_i \cdots g_N}$. Within this subspace, the symmetric collective state $\ket{W_{\mathrm{Ryd}}} = N^{-1/2}\sum_{i=1}^N\ket{r_i}$ is the only state coupled to $\ket{G_e}$, with collectively enhanced Rabi frequency $\Omega_{\mathrm{eff}} = \sqrt{N}\,\Omega$, reducing the ensemble to an effective two-level system.

The dynamics of the system are described within the four-state basis $\{\ket{g_c,G_e},\,\ket{r_c,G_e},\,\ket{g_c,W_{\mathrm{Ryd}}},\, \ket{r_c,W_{\mathrm{Ryd}}}\}$. Working in the rotating frame at zero detuning, the Hamiltonian is
\begin{equation}
  H = H_c
    + \frac{\hbar\Omega_{\mathrm{eff}}}{2}
      \bigl(\ket{W_{\mathrm{Ryd}}}\!\bra{G_e} + \mathrm{h.c.}\bigr)
    + V\ket{r_c,W_{\mathrm{Ryd}}}\!\bra{r_c,W_{\mathrm{Ryd}}},
  \label{eq:hamiltonian}
\end{equation}
where $H_c = \frac{\hbar\Omega}{2}(\ket{r_c}\!\bra{g_c} + \mathrm{h.c.})$ drives the control atom, the second term is the collectively enhanced ensemble transition, and $V = C_6/r^6$ is the van der Waals interaction energy between the control atom and the ensemble excitation at their separation $r$. In the strong blockade regime $V \gg \Omega_{\mathrm{eff}}$, the doubly-excited state $\ket{r_c,W_{\mathrm{Ryd}}}$ is shifted far off resonance, restricting the dynamics to the three-state subspace $\{\ket{g_c,G_e},\,\ket{r_c,G_e},\,\ket{g_c,W_{\mathrm{Ryd}}}\}$ (derivation in SM Sec.~S1).

Entanglement is generated through a two-pulse sequence initialized in $\ket{g_c,G_e}$. The first pulse addresses the control atom with driving laser phase $\phi_L=\pi/2$, implementing a $R_y(\pi/2)$ rotation that prepares a real equal-weight superposition, $R_y(\pi/2)\ket{g_c} = (\ket{g_c}+\ket{r_c})/\sqrt{2}$, without introducing any relative phase. After the first pulse the composite system is therefore
\begin{equation}
  \ket{\psi_c}= \frac{1}{\sqrt{2}} \bigl( \ket{g_c,G_e} + \ket{r_c,G_e} \bigr), \label{eq:pi2pulse}
\end{equation}
with the ensemble remaining in $\ket{G_e}$. The second pulse resonantly drives the ensemble transition $\ket{G_e}\leftrightarrow\ket{W_{\mathrm{Ryd}}}$ under $H_e = (\hbar\Omega_{\mathrm{eff}}/2) (\ket{W_{\mathrm{Ryd}}}\bra{G_e}+\mathrm{h.c.})$ for a duration corresponding to a collective $\pi$ pulse. A resonant $\pi$ pulse gives $R_x(\pi)\ket{G_e} = -i\ket{W_{\mathrm{Ryd}}}$, where the $-i$ is the standard dynamical phase of a resonant $\pi$ rotation. The Rydberg interaction makes this operation conditional on the control-atom state: if the control atom is in $\ket{g_c}$, the ensemble undergoes the full $\pi$ rotation; if it is in $\ket{r_c}$, the doubly-excited state $\ket{r_c,W_{\mathrm{Ryd}}}$ is shifted out of resonance by $V$, suppressing the ensemble excitation in the blockade regime $V\gg\hbar\Omega_{\mathrm{eff}}$. To leading order in $\hbar\Omega_{\mathrm{eff}}/V$, the conditional evolution is
\begin{equation}
\ket{g_c,G_e} \rightarrow -i\ket{g_c,W_{\mathrm{Ryd}}}, \qquad \ket{r_c,G_e} \rightarrow \ket{r_c,G_e}.
  \label{eq:conditional}
\end{equation}
Applying this to Eq.~\eqref{eq:pi2pulse} yields the maximally entangled Bell state
\begin{equation}
\ket{\psi_1} = \frac{1}{\sqrt{2}} \bigl( \ket{r_c,G_e} - i\ket{g_c,W_{\mathrm{Ryd}}} \bigr),
  \label{eq:bell_state}
\end{equation}
correlating the Rydberg occupation of the control atom with the collective ensemble excitation. The relative phase $-i$ is set entirely by the ensemble $\pi$ pulse (SM Sec.~S1).

\begin{figure}[t]
\centering
\includegraphics[width=\linewidth,keepaspectratio]{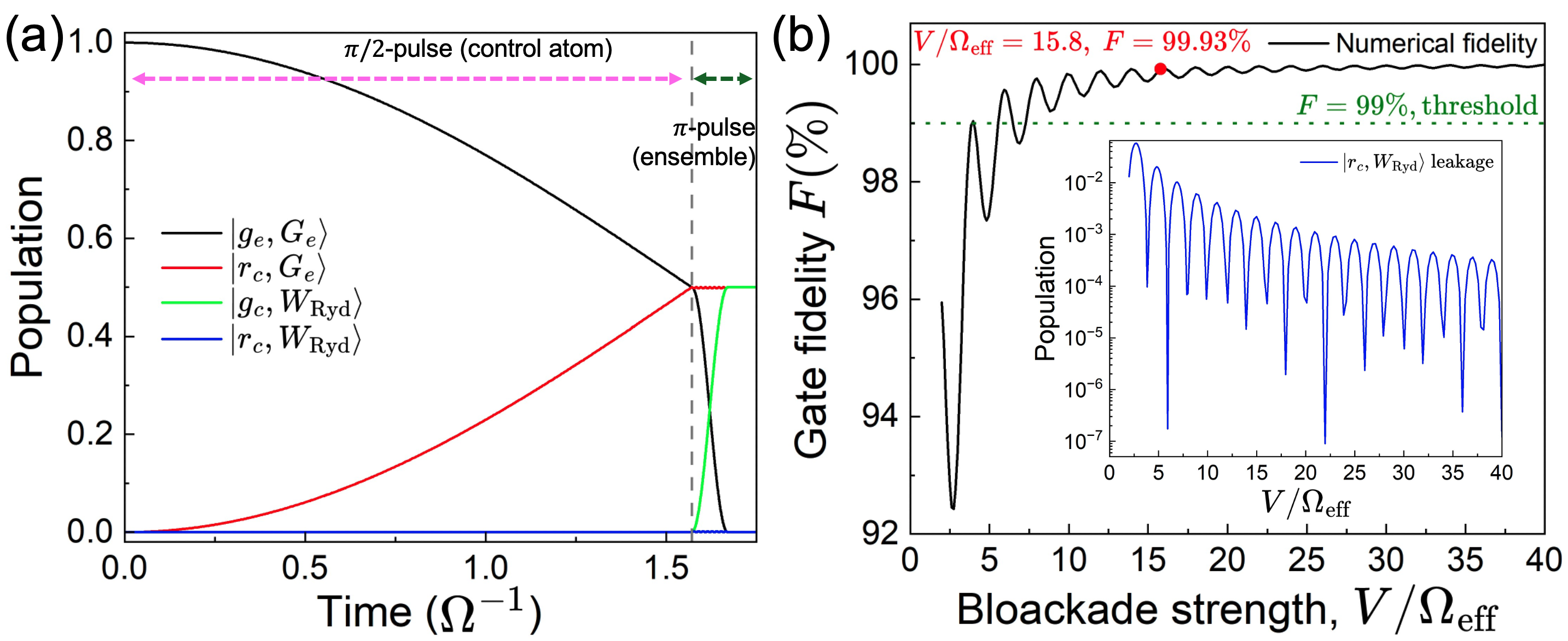}
\caption{%
  \textbf{Bell-state generation via Rydberg blockade.}
  \textbf{(a)}~Population dynamics during the two-pulse protocol ($N = 1000$, $V/\Omega_{\mathrm{eff}} = 15.8$). A $\pi/2$ pulse on the control atom (magenta arrow) followed by a conditional $\pi$ pulse on the ensemble (green arrow) generates the Bell state $\ket{\psi_1}$. The doubly-excited state $\ket{r_c,W_{\mathrm{Ryd}}}$ (blue) remains below $10^{-4}$ throughout, confirming single-excitation confinement.
  \textbf{(b)}~Gate fidelity $F$ as a function of blockade strength $V/\Omega_{\mathrm{eff}}$, computed numerically. The only infidelity source is leakage into $\ket{r_c,W_{\mathrm{Ryd}}}$ (inset), which is rapidly suppressed with increasing blockade strength. Fidelity exceeds 99\% for $V/\Omega_{\mathrm{eff}} \gtrsim 10$. The operating point (red dot, $V/\Omega_{\mathrm{eff}} = 15.8$, $F = 99.93\%$) lies well within the high-fidelity regime.}
\label{fig:step1}
\end{figure}

Figure~\ref{fig:step1}a shows the population dynamics obtained by numerically integrating Eq.~\eqref{eq:hamiltonian} with the two pulses applied sequentially. The Bell state $\ket{\psi_1}$ is generated with fidelity $F = 99.93\%$ for $N = 1000$ atoms at blockade strength $V/\Omega_{\mathrm{eff}} = 15.8$. Collective enhancement enables fast operation: for $\Omega/2\pi = 10$~MHz, the effective coupling reaches $\Omega_{\mathrm{eff}}/2\pi \approx 316$~MHz, corresponding to a $\pi$-pulse duration of $t_\pi \approx 1.6$~ns. The population of $\ket{r_c,W_{\mathrm{Ryd}}}$ (blue curve) remains below $10^{-4}$ throughout, confirming that the blockade confines all dynamics to the single-excitation manifold.

Figure~\ref{fig:step1}b shows the gate fidelity as a function of blockade strength $V/\Omega_{\mathrm{eff}}$. The sole source of infidelity in this model is leakage into $\ket{r_c,W_{\mathrm{Ryd}}}$, which falls rapidly with increasing $V/\Omega_{\mathrm{eff}}$, driving $F$ above $99\%$ for $V/\Omega_{\mathrm{eff}} \gtrsim 10$. The operating point $V/\Omega_{\mathrm{eff}} = 15.8$ balances fidelity against experimental accessibility: it delivers $F = 99.93\%$ while requiring a blockade strength $V/\Omega = 15.8\times\sqrt{N} \approx 500$, achievable with $^{87}$Rb Rydberg states at $n \sim 60$--70 and cold-atom densities $\rho \sim 10^{10}$--$10^{11}$~cm$^{-3}$. Larger ratios would demand higher principal quantum numbers with correspondingly shorter Rydberg lifetimes, and the fidelity curve saturates beyond this point. Full derivations of the Hamiltonian reduction, blockade leakage scaling, and achievable parameter ranges are in SM Sec.~S1.

At larger ensemble sizes, atom-number fluctuations ($\delta N \sim \sqrt{N}$) and multi-level leakage into off-resonant magnetic sublevels reduce the fidelity beyond what is captured by Eq.~\eqref{eq:hamiltonian}. These effects can be suppressed using DRAG pulses~\cite{motzoi2009simple,gambetta2011analytic} and a bias magnetic field that spectrally resolves the target transition from neighbouring sublevels~\cite{lancaster2025quantum}. The operating point $N = 1000$ balances collective enhancement against these error sources. The Bell state $\ket{\psi_1}$ serves as the input resource for Step~2, where $\ket{W_{\mathrm{Ryd}}}$ is mapped onto a photonic qubit.

\section{Atom--Photon Interface}\label{atom_photon}
An efficient atom--photon interface for remote quantum communication requires three capabilities: (i)~coherent conversion of $\ket{W_{\mathrm{Ryd}}}$ into a propagating photon, (ii)~directional emission for efficient photon coupling, and (iii)~encoding of the quantum information in a photonic degree of freedom robust against fiber decoherence. All three are achieved by two-photon Raman pulse.

The two-photon Raman Hamiltonian
\begin{equation}
  H_{\mathrm{Raman}}
  = \frac{\hbar\Omega_R}{2}
    \sum_{i=1}^{N} e^{i\mathbf{k}_0\cdot\mathbf{r}_i}
    \bigl(\ket{e_i,m_j'}\!\bra{r_i,m_j}+\mathrm{h.c.}\bigr)
  \label{eq:raman_hamiltonian}
\end{equation}
drives each atom from its Rydberg sublevel $m_j$ to the optically excited sublevel $m_j'$ (e.g.\ $5P_{3/2}$ in $^{87}$Rb), where $\Omega_R = \Omega_1\Omega_2/(2\Delta)$ is the two-photon Rabi frequency ($\Omega_{1,2}$: single-photon Rabi frequencies, $\Delta$: intermediate-state detuning), and $\mathbf{k}_0 = \mathbf{k}_1 - \mathbf{k}_2$ is the wavevector difference of the two Raman beams, which defines the target emission direction $\hat{\mathbf{k}}_0$.

Acting on $\ket{W_{\mathrm{Ryd}}} = N^{-1/2}\sum_i\ket{r_i}$, $H_{\mathrm{Raman}}$ transfers the collective excitation to the optically active state
\begin{equation}
  \ket{W_e}
  = \frac{1}{\sqrt{N}}\sum_{i=1}^{N}
    e^{i\mathbf{k}_0\cdot\mathbf{r}_i}\ket{e_i}.
  \label{eq:optical_collective}
\end{equation}
The position-dependent phase $e^{i\mathbf{k}_0\cdot\mathbf{r}_i}$ encodes the target emission direction into the spatial coherence of the state and is preserved through the spontaneous emission process. Because $\ket{W_e}$ inherits the $\sqrt{N}$ collective enhancement, the emission rate is $\Gamma_{\mathrm{coll}} = N\Gamma_0$ \cite{gross1982superradiance,dicke1954coherence,masson2020many}, giving $\tau_{\mathrm{emit}} \approx 160$~ps for $N=1000$ and $\Gamma_0/2\pi \approx 6$~MHz.

\begin{figure*}[t]
\centering
\includegraphics[width=\textwidth,keepaspectratio]{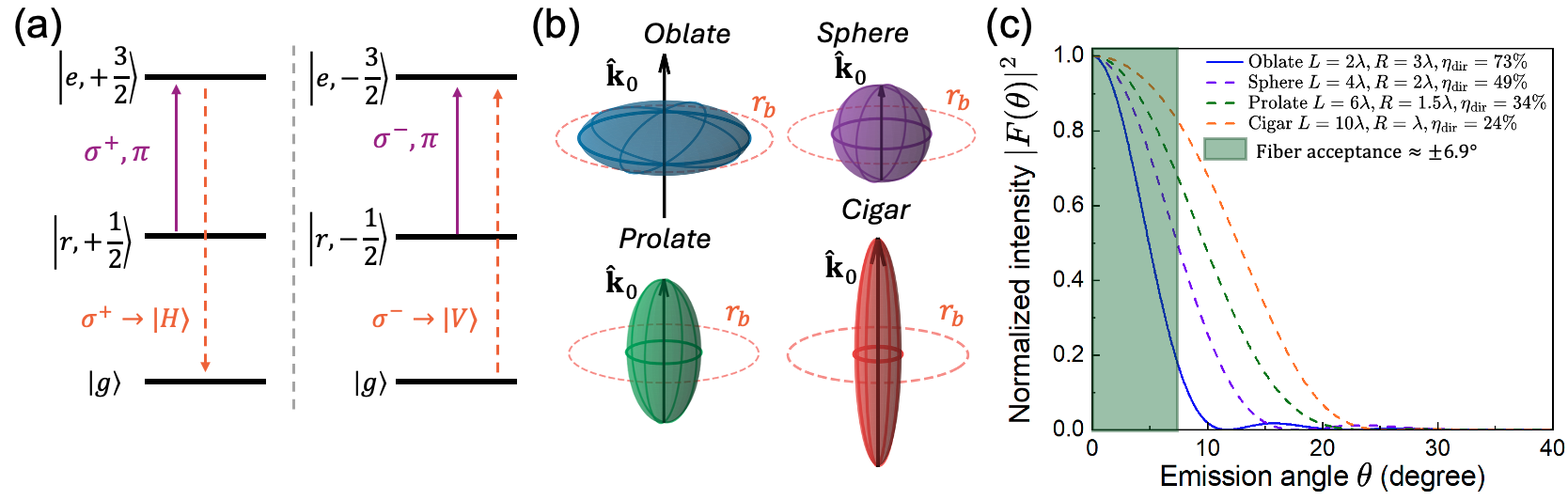}
\caption{%
  \textbf{Atom--photon interface.}
  \textbf{(a)}~State-selective Raman transitions for polarization
  encoding.
  Beams $(\sigma^+,\pi)$ drive $\ket{r,+\tfrac{1}{2}} \to
  \ket{e,+\tfrac{3}{2}}$ (left) and $(\sigma^-,\pi)$ drive
  $\ket{r,-\tfrac{1}{2}} \to \ket{e,-\tfrac{3}{2}}$ (right);
  the emitted $\sigma^\pm$ photons are converted to $\ket{H}$,
  $\ket{V}$ by a quarter-wave plate.
  \textbf{(b)}~The four cloud geometries classified by the ratio
  of equatorial semi-axis $R$ to polar semi-axis $L/2$:
  oblate ($R>L/2$, blue), sphere ($R=L/2$, purple),
  prolate ($R<L/2$, green), and cigar ($R\ll L/2$, red;
  violates blockade constraint, shown for reference).
  The dashed circle marks the blockade radius $r_b = 3.8\lambda$;
  $\hat{\mathbf{k}}_0$ indicates the emission direction.
  \textbf{(c)}~Normalized angular emission pattern $|F(\theta)|^2$
  (Eq.~\eqref{eq:array_factor}) for all four geometries.
  The green shading marks the fiber acceptance cone
  ($\pm6.9^\circ$, NA$=0.12$).
  $\eta_{\mathrm{dir}}$ increases monotonically with $R$:
  $73\%$ (oblate), $49\%$ (sphere), $34\%$ (prolate),
  $24\%$ (cigar, reference only).
}
\label{fig:step2}
\end{figure*}

Polarization encoding exploits the Zeeman structure of the Rydberg state. The two Raman beam polarizations $(\sigma^\pm,\,\pi)$ couple the sublevels $m_j = \pm 1/2$ to distinct excited states through net angular momentum transfers $\Delta m = \pm 1$ (see figure~\ref{fig:step2}a):
\begin{equation}
\begin{aligned}
  \ket{r,+\tfrac{1}{2}}
  &\;\xrightarrow{\;\sigma^+,\pi\;}\;
  \ket{e,+\tfrac{3}{2}}
  \;\xrightarrow{\;\mathrm{decay}\;}\;
  \ket{g,+\tfrac{1}{2}} + \ket{H}, \\[4pt]
  \ket{r,-\tfrac{1}{2}}
  &\;\xrightarrow{\;\sigma^-,\pi\;}\;
  \ket{e,-\tfrac{3}{2}}
  \;\xrightarrow{\;\mathrm{decay}\;}\;
  \ket{g,-\tfrac{1}{2}} + \ket{V}.
\end{aligned}
\label{eq:polarization_selective}
\end{equation}
Each excited state decays under the selection rule $m_{\mathrm{photon}} = m_e - m_g$, emitting $\sigma^+$ from $\ket{e,+\tfrac{3}{2}}$ and $\sigma^-$ from $\ket{e,-\tfrac{3}{2}}$; a quarter-wave plate then converts $\sigma^\pm$ to $\{\ket{H},\ket{V}\}$. Within the fiber acceptance cone ($\Delta\theta < 10^\circ$), $\sigma$--$\pi$ mixing is negligible and the in-cone polarization purity exceeds $95\%$ (see details in SM section~S2).

Since the same $H_{\mathrm{Raman}}$ drives both pathways coherently, initializing the control atom in an equal superposition of both Zeeman sublevels before Step~2 produces, starting from the Bell state $\ket{\psi_1} = (\ket{r_c,G_e} - i\ket{g_c,W_{\mathrm{Ryd}}})/\sqrt{2}$ where $\ket{G_e} = \bigotimes_{i=1}^{N} \ket{g_i}$ is the collective ground state, the polarization-entangled atom--photon state
\begin{equation}
  \ket{\psi_2}
  = \frac{1}{\sqrt{2}}
    \bigl(
      \ket{r_c,+\tfrac{1}{2}}\ket{G_e}\ket{H}
     +\ket{r_c,-\tfrac{1}{2}}\ket{G_e}\ket{V}
    \bigr).
  \label{eq:psi2}
\end{equation}
The Zeeman spin of the control atom is thus entangled with the polarization of a photon emitted along $\hat{\mathbf{k}}_0$; $\ket{\psi_2}$ is the essential resource for the remote entanglement protocol of Step~3.

While $\ket{\psi_2}$ constitutes the quantum resource for Step~3, the rate at which it is generated depends on how efficiently the emitted photon couples into the fiber mode — a quantity governed entirely by the geometry of the atomic ensemble. The phase imprinting $e^{i\mathbf{k}_0\cdot\mathbf{r}_i}$ in $\ket{W_e}$ creates directional emission through quantum interference: photons emitted along $\hat{\mathbf{k}}_0$ from all $N$ atoms add constructively, while emission into other directions is suppressed.

The emission probability into direction $\theta$ (polar angle from $\hat{\mathbf{k}}_0$) is proportional to $|F(\theta)|^2$, where $F(\theta) = N^{-1}\sum_i e^{i(\mathbf{k}_0-\mathbf{k})\cdot\mathbf{r}_i}$ is the normalised array factor and $k_0 = 2\pi/\lambda$. For a cylindrical cloud of length $L$ along $\hat{\mathbf{k}}_0$ and equatorial radius $R$, $F(\theta)$ factorises exactly as (SM section~S3)
\begin{equation}
  F(\theta) =
  \underbrace{
    \mathrm{sinc}\!\left(\frac{k_0 L(1-\cos\theta)}{2}\right)
  }_{F_{\mathrm{long}}(\theta)}
  \times
  \underbrace{
    \frac{2J_1(k_0 R\sin\theta)}{k_0 R\sin\theta}
  }_{F_{\mathrm{trans}}(\theta)},
  \label{eq:array_factor}
\end{equation}
where $\mathrm{sinc}(x) = \sin(x)/x$ and $J_1$ is the first-order Bessel function of the first kind. $F_{\mathrm{long}}$ concentrates emission within a cone of half-angle $\sqrt{2\lambda/L}$, while $F_{\mathrm{trans}}$ sets a cone of half-angle $0.61\lambda/R$; the tighter of the two governs the beam width. Because the longitudinal mismatch $\Delta k_z = k_0(1-\cos\theta)$ grows as $\theta^2$ at small angles while the transverse mismatch $\Delta k_\perp = k_0\sin\theta$ grows linearly, $F_{\mathrm{trans}}$ imposes the tighter cone whenever $0.61\lambda/R < \sqrt{2\lambda/L}$, which holds for all three blockade-consistent geometries considered here and reduces the beam width to a function of $R$ alone.

The geometry is constrained by the blockade condition of Step~1: all atoms must lie within the blockade radius $r_b \approx 3.8\lambda$ of the control atom ($V/\Omega_{\mathrm{eff}} = 15.8$), giving $r_{\max} = \sqrt{R^2+(L/2)^2} \leq r_b$. The directional efficiency
\begin{equation}
  \eta_{\mathrm{dir}}
  = \frac{\displaystyle\int_0^{\theta_{\max}}
          |F(\theta)|^2\sin\theta\,d\theta}
         {\displaystyle\int_0^{\pi}
          |F(\theta)|^2\sin\theta\,d\theta},
  \label{eq:eta_dir}
\end{equation}
the fraction of emitted photons collected by the fiber ($\theta_{\max} = 6.9^\circ$ for NA$=0.12$, Thorlabs P3-780A), increases monotonically with $R$, so the optimal geometry maximises $R$ within the blockade constraint. The four canonical shapes — oblate ($R>L/2$), sphere ($R=L/2$), prolate ($R<L/2$), and cigar ($R\ll L/2$) — are illustrated in figure~\ref{fig:step2}b and span the full design space. The resulting efficiencies, shown in Fig.~\ref{fig:step2}c, are $73\%$ for the oblate spheroid ($L=2\lambda$, $R=3\lambda$, $r_{\max}=3.2\lambda$), $49\%$ for the sphere ($L=4\lambda$, $R=2\lambda$), and $34\%$ for the prolate spheroid ($L=6\lambda$, $R=1.5\lambda$). The cigar geometry ($L=10\lambda$, $R=\lambda$, $r_{\max}=5.1\lambda>r_b$) violates the blockade constraint which makes this shape unsuitable for directional emission. The oblate spheroid, which achieves the highest $\eta_{\mathrm{dir}}$ among all blockade-consistent geometries, is adopted throughout this work. The full derivations for all four cases are provided in in SM section~S3.

\section{Remote Entanglement via Photonic Interference}
\label{sec:step3}
Remote entanglement between two spatially separated nodes is generated by interfering the photons emitted in Step~2 at a central beam-splitter station and conditioning on coincidence detection, as illustrated in Fig.~\ref{fig:protocol_overview}c. For the two identical nodes, labeled as $A$ and $B$, each node emits one photon and is prepared in the atom--photon entangled state $\ket{\psi_2}$ (Eq.~\eqref{eq:psi2}). For compactness, we define, $\ket{+}\equiv\ket{r,m_j=+\tfrac{1}{2}}$ and $\ket{-}\equiv\ket{r,m_j=-\tfrac{1}{2}}$, and suppress the collective ground state $\ket{G_e}$ (appears identically in all terms and does not affect the entanglement structure). The combined atom--photon state of the two nodes is then
\begin{align}
&\ket{\Psi_{AB}}
= \frac{1}{2}\Bigl(
\ket{+}_A\ket{+}_B \ket{H}_A\ket{H}_B
+\ket{+}_A\ket{-}_B \ket{H}_A\ket{V}_B
\nonumber\\
&+\ket{-}_A\ket{+}_B \ket{V}_A\ket{H}_B
+\ket{-}_A\ket{-}_B \ket{V}_A\ket{V}_B
\Bigr).
\label{eq:Psi_AB}
\end{align}
The photons from nodes $A$ and $B$ are directed to the two input ports ($a$ and $b$) of a 50:50 non-polarizing beamsplitter. By Hong--Ou--Mandel interference~\cite{hong1987measurement,duan2001long}, photon pairs with identical polarizations ($\ket{H}_A\ket{H}_B$ or $\ket{V}_A\ket{V}_B$) gather into the same output port and do not contribute to coincidence events. Cross-polarized pairs remain distinguishable and can produce one photon in each output port. The two output ports ($c$ and $d$) are each monitored by a polarization-resolving single-photon detector. The full beamsplitter transformation and post-selection are detailed in SM section~S4. Applying the beamsplitter transformation to Eq.~\eqref{eq:Psi_AB} and retaining only events in which one photon is detected in each output port with orthogonal polarizations yields the two heralded atomic states
\begin{align}
\ket{\psi_{H_c,V_d}}
&= \frac{1}{\sqrt{2}}\bigl(-\ket{+}_A\ket{-}_B + \ket{-}_A\ket{+}_B\bigr),
\label{eq:herald1}\\[4pt]
\ket{\psi_{V_c,H_d}}
&= \frac{1}{\sqrt{2}}\bigl(\ket{+}_A\ket{-}_B - \ket{-}_A\ket{+}_B\bigr).
\label{eq:herald2}
\end{align}
Both outcomes are equivalent to the maximally entangled Bell state $\ket{\Psi^-}_{AB} = (\ket{+}_A\ket{-}_B - \ket{-}_A\ket{+}_B)/\sqrt{2}$ up to an overall phase, so both herald the same entangled state regardless of the coincidence outcome. The probability of a coincidence event per protocol trial is $P_{\mathrm{success}} = 1/4$, arising because only the two cross-polarized terms in Eq.~\eqref{eq:Psi_AB} contribute to coincidences (factor $1/2$), and each contributes with amplitude $1/2$ to each output combination. A coincidence detection event at the central station therefore heralds the deterministic generation of the Bell state $\ket{\Psi^-}$ between the control atoms of the two remote nodes.

\section{Protocol Efficiency}\label{prot-eff}
The efficiency of a single node is the product of the gate fidelity,
directional collection efficiency, and Raman mapping efficiency:
\begin{equation}
  \eta_{\mathrm{node}}
  = F_{\mathrm{gate}} \times \eta_{\mathrm{dir}} \times \eta_{\mathrm{map}}.
  \label{eq:eta_node}
\end{equation}
Here $F_{\mathrm{gate}} \approx 0.99$ is the Bell-state fidelity from
Step~1 (Sec.~\ref{local-entanglement}); $\eta_{\mathrm{dir}} = 73\%$
is the directional efficiency for the oblate spheroid geometry
($L=2\lambda$, $R=3\lambda$, Sec.~\ref{atom_photon}); and
$\eta_{\mathrm{map}} \approx 0.55$ is the Raman mapping efficiency,
which decomposes as
$\eta_{\mathrm{map}} = \eta_{\mathrm{Raman}} \times \eta_{\mathrm{SMF}}
\times \eta_{\mathrm{pol}} \approx 0.80 \times 0.80 \times 0.86$
(SM Sec.~S5.7), accounting for Raman transfer losses, fiber
mode-matching, and polarization purity respectively.
Together, the single-node efficiency is
$\eta_{\mathrm{node}} = 0.99 \times 0.73 \times 0.55 \approx 0.40$,
giving the probability that a successful Step-1 operation produces a
high-fidelity atom--photon entangled state with the photon coupled into
the target fiber mode.

The success probability per attempt is
\begin{equation}
  P_E
  = \frac{1}{4}\,
    \eta_{\mathrm{node}}^2\,
    \eta_{\mathrm{prop}}^2\,
    \eta_{\mathrm{det}}^2,
  \label{eq:PE}
\end{equation}
where the factor $1/4$ is the heralding probability from beamsplitter
post-selection (Sec.~\ref{sec:step3}).
The fiber propagation efficiency is
$\eta_{\mathrm{prop}}(L) = 10^{-\alpha L/10}$, with
$\alpha = 0.2$~dB/km the attenuation of standard telecom single-mode
fiber at $\sim$1550~nm~\cite{van2022entangling}; this value applies
after quantum frequency conversion from the native 780~nm photons,
since attenuation at 780~nm is $\sim$2--3~dB/km.
Note that fiber mode-matching is already included in $\eta_{\mathrm{map}}$
via $\eta_{\mathrm{SMF}}$, so $\eta_{\mathrm{prop}}$ represents fiber
attenuation only.
The detector efficiency $\eta_{\mathrm{det}} \approx 0.80$ is a
conservative estimate for superconducting nanowire single-photon
detectors (SNSPDs); the reference devices
achieve $\eta_{\mathrm{det}} = 93\%$~\cite{marsili2013detecting} and
recent state-of-the-art SNSPDs exceed $99\%$~\cite{li2025surpassing},
which would increase $P_E$ by factors of $1.35$ and $1.53$
respectively.
At metropolitan scale ($L = 20$~km),
$\eta_{\mathrm{prop}} = 10^{-0.4} \approx 0.40$, giving
$P_E \approx 0.40\%$ with the conservative $\eta_{\mathrm{det}} = 0.80$.

\begin{figure}[h]
\centering
\includegraphics[width=\linewidth,keepaspectratio]{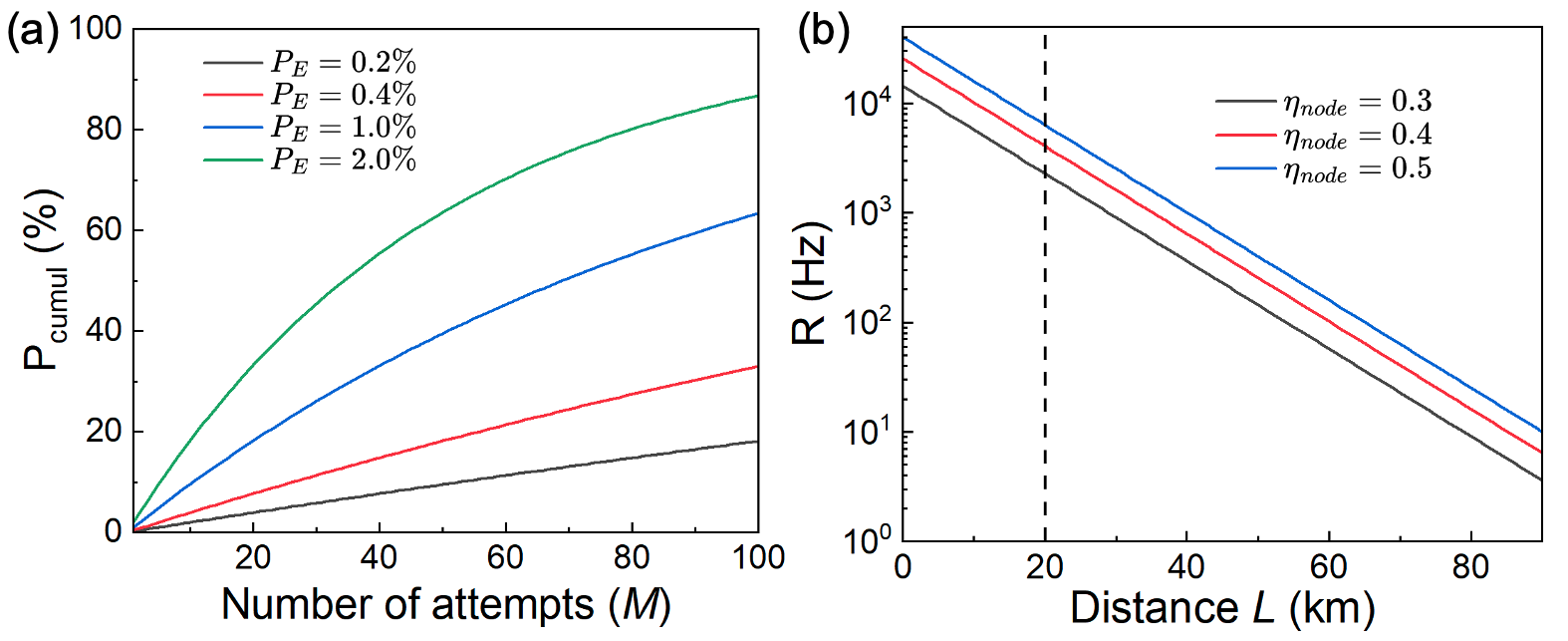}
\caption{%
  \textbf{Protocol efficiency.}
  \textbf{(a)}~Cumulative success probability ($P_{\mathrm{cumul}}(M)$) vs.\ number of attempts $M$ for
  $P_E = 0.2\%$ (black), $0.4\%$ (red), $1.0\%$ (blue), and $2.0\%$ (green). The red curve ($P_E = 0.40\%$) corresponds to the operating point at 20~km; $M = 100$ attempts yield $P_{\mathrm{cumul}} \approx 33\%$.
  \textbf{(b)}~Entanglement generation rate $R$ (R(L)) vs.\ distance for $\eta_{\mathrm{node}} = 0.30$ (black), $0.40$ (red), and $0.50$ (blue). Dashed line marks the 20 km. Parameters: $\alpha = 0.2$~dB/km, $\eta_{\mathrm{det}} = 0.80$, $T_{\mathrm{cycle}} = 1\,\mu$s.}
\label{fig:efficiency}
\end{figure}

Since $P_E \ll 1$, an electromagnetically induced transparency (EIT)
quantum memory (SM Sec.~S5.1) stores the atom--photon entangled state
between successive remote-entanglement attempts.
With coherence time $T_2 > 100\,\mu$s and cycle time
$T_{\mathrm{cycle}} \approx 1\,\mu$s, up to
$M_{\mathrm{max}} \approx 100$ attempts are possible within $T_2$.
The cumulative success probability after $M$ attempts is
\begin{equation}
  P_{\mathrm{cumul}}(M) = 1-(1-P_E)^M.
  \label{eq:Pcumul}
\end{equation}
For $P_E = 0.40\%$ and $M=100$,
$P_{\mathrm{cumul}} \approx 33\%$ (Fig.~\ref{fig:efficiency}a).

The entanglement generation rate is $R = P_E/T_{\mathrm{cycle}}$.
Substituting $\eta_{\mathrm{prop}}(L) = 10^{-\alpha L/10}$:
\begin{equation}
  R(L) = \frac{\eta_{\mathrm{node}}^2\,\eta_{\mathrm{det}}^2}
              {4\,T_{\mathrm{cycle}}}
         \times 10^{-\alpha L/5},
  \label{eq:RL}
\end{equation}
with $L$ in km.
For $\eta_{\mathrm{node}} = 0.40$, Eq.~\eqref{eq:RL} gives
$R \approx 4\,\mathrm{kHz}$ at 20~km (Fig.~\ref{fig:efficiency}b),
rising to $\approx 6\,\mathrm{kHz}$ for $\eta_{\mathrm{node}} = 0.50$.
Upgrading to state-of-the-art SNSPDs with $\eta_{\mathrm{det}} = 0.99$
would further increase the rate to $\approx 6\,\mathrm{kHz}$ at
$\eta_{\mathrm{node}} = 0.40$, providing a clear near-term improvement
pathway.

\section{Discussion and Conclusion}
We have presented a cavity-free protocol for distributed quantum computing that employs Rydberg atom ensembles to integrate high-fidelity local entanglement generation, efficient atom--photon conversion, and remote entanglement distribution over
metropolitan-scale distances. The protocol achieves Bell-state fidelities exceeding $97.5\%$, a node efficiency $\eta_{\mathrm{node}} \approx 40\%$, and entanglement generation rates of $\approx 4\,\mathrm{kHz}$ at 20~km separation. These results demonstrate that practical quantum networks can be implemented without optical cavities, offering a viable platform for distributed quantum computing.

The central advantage of the protocol is the collective enhancement inherent in the Rydberg ensemble, which enables fast local gates and directional photon emission without the fabrication and mode-matching constraints of cavity-based approaches. The oblate spheroid geometry ($L=2\lambda$, $R=3\lambda$), which maximises the equatorial radius $R$ within the blockade constraint $r_{\max} \leq r_b = 3.8\lambda$, achieves a directional efficiency $\eta_{\mathrm{dir}} = 73\%$, comparable to moderate-finesse optical cavities. Combined with the Raman mapping efficiency $\eta_{\mathrm{map}} \approx 0.55$ and gate fidelity $F_{\mathrm{gate}} \approx 0.99$, the resulting $\eta_{\mathrm{node}} \approx 0.40$ places the protocol firmly in the regime where quantum memory-enhanced repeated attempts ($M \lesssim 100$ within $T_2 > 100\,\mu$s) yield cumulative entanglement probabilities approaching $33\%$ per coherence window at metropolitan scale.

The protocol is compatible with existing Rydberg platforms~\cite{saffman2010quantum,levine2019parallel,covey2023quantum}. The required laser systems (780~nm for optical transitions, 480~nm for Rydberg excitation) are commercially available. Detector efficiencies $\eta_{\mathrm{det}} \approx 0.80$ are representative of current superconducting nanowire single-photon detectors~\cite{marsili2013detecting,li2025surpassing}; state-of-the-art devices achieve $\eta_{\mathrm{det}} > 0.95$, which would increase the entanglement rate by a further factor of $\sim1.4$. Recent demonstrations of atom--photon entanglement over 20~km~\cite{van2020long} and atom--atom entanglement over 33~km~\cite{van2022entangling} in deployed fiber networks confirm the feasibility of the proposed operational range.

Experimental implementation requires careful control of several decoherence and loss mechanisms, including leakage to nearby Rydberg states in the multilevel atomic structure~\cite{mulliken1976rydberg}, dephasing due to atomic position fluctuations~\cite{jaksch2000fast,
mewes2005decoherence,jiang2016dynamical}, Raman phase noise, photon reabsorption, fiber transmission losses, and polarization instabilities~\cite{sasaki1987long,galtarossa1994polarization}. These can be mitigated through optimised laser detunings, magnetic-field control, and tapered pulse shaping~\cite{muller2011prospects,goerz2014robustness}.

In summary, the proposed scheme provides a practical, experimentally accessible framework for scalable quantum networking without optical cavities. The dominant remaining bottleneck is the Raman mapping efficiency $\eta_{\mathrm{map}} \approx 0.55$; improvements in fiber mode-matching ($\eta_{\mathrm{SMF}}$) and Raman transfer ($\eta_{\mathrm{Raman}}$) each at the $10\%$ level would raise $\eta_{\mathrm{node}}$ to $\sim0.55$, pushing the 20~km rate above $7\,\mathrm{kHz}$. Long-distance operation ($L > 50\,\mathrm{km}$) will additionally require quantum frequency conversion from 780~nm to the telecom band. Future work will focus on experimental realization, multi-node network architectures, and integration with quantum error correction schemes for fault-tolerant distributed quantum computing.

\section{Data availability statement}
All data that support the findings of this study are included within the article.

\section{Conflict of interest statement}
The authors declare no conflict of interest.

\section{Author Contributions}

A.U. conceived the project, developed 
the theoretical framework, performed all analytical and numerical 
calculations, produced the figures, and wrote the original 
manuscript. M.A.S. provided experimental insight and 
contributed to manuscript review and editing.
All authors discussed the results and approved the final manuscript.


\bibliography{ref}
\end{document}